\documentclass{aa}  

\usepackage{graphics,graphicx}
\usepackage{xcolor}
\usepackage{enumitem}
\usepackage{caption}

\usepackage{txfonts}
\usepackage{longtable}
\usepackage[unicode=true,pdfusetitle,
 bookmarks=true,bookmarksnumbered=false,bookmarksopen=false,
 breaklinks=false,pdfborder={0 0 0},pdfborderstyle={},backref=false,colorlinks=true]
 {hyperref}
\hypersetup{
 urlcolor=blue,citecolor=blue}

\begin{document}

   \title{Probing habitable regions with SRG/eROSITA}

   \author{E. Gatuzz\inst{1},   
           S. Rukdee\inst{1}, 
           S. Freund\inst{1} and
           T. Kallman\inst{2}   
          }

   \institute{Max-Planck-Institut f\"ur extraterrestrische Physik, Gie{\ss}enbachstra{\ss}e 1, 85748 Garching, Germany\\
              \email{egatuzz@mpe.mpg.de} 
             \and
     NASA / Goddard Space Flight Center, Greenbelt, MD 20771, USA 
             }

   \date{Received XXX; accepted YYY}

  \abstract  
{
Stellar high-energy radiation is a key driver of atmospheric erosion and evolution in exoplanets, directly affecting their long-term habitability.
We present a comprehensive study on stellar high-energy radiation and its impact on exoplanetary atmospheres, leveraging data from the \textit{SRG/eROSITA} all-sky survey. 
Our sample consists of 3750 main-sequence stars identified by cross-matching with \textit{Gaia} DR3. 
Utilizing X-ray spectral fits from the \textit{eROSITA} catalog, we computed X-ray ($L_X$) and combined extreme-ultraviolet (EUV) luminosities ($L_{\mathrm{EUV}}$), which we used to derive XUV fluxes at the habitable zone ($F_{\mathrm{XUV,HZ}}$). 
We find that the majority of stars in our sample are significantly more XUV-active than the Sun, with habitable zone fluxes ranging from $10^0$ to $10^5$ erg~cm$^{-2}$~s$^{-1}$. 
The ratio of $L_{\mathrm{XUV}}/L_{\mathrm{bol}}$ is found to be higher for cooler, magnetically active stars, highlighting their potentially hazardous nature for planetary atmospheres. 
Applying the energy-limited escape model, we computed atmospheric mass-loss rates for hypothetical earth-like planets located at the habitable zone of each star. 
We also present local maps for distances up to $500$~pc of the average XUV flux, revealing ``hazard zones'' where stellar radiation could significantly influence planetary atmospheric evolution. 
This work demonstrates the power of X-ray surveys in constraining the high-energy environments of exoplanets and underscores the critical role of stellar activity in planetary habitability.  
}
 
    \titlerunning{Probing habitable regions with SRG/eROSITA}
    \authorrunning{Gatuzz et al.}
   \maketitle

\section{Introduction} 
Exoplanets have been discovered in a wide range of orbital configurations, with a notable fraction orbiting close to their host stars. 
Such close-in planets, including the first main-sequence exoplanet 51 Peg b \citep{may95}, experience stellar irradiation that can exceed the levels received by planets in our Solar System by orders of magnitude. 
This high-energy irradiation, spanning from the ultraviolet (UV) to X-rays, can cause significant atmospheric heating, planetary radius inflation \citep{for10, bar10}, and even drive hydrodynamic escape \citep{wat81, mur09}.

Observational evidence for extended planetary atmospheres and ongoing atmospheric escape has been obtained through transit spectroscopy across various spectral bands. 
Notable diagnostic features include the Lyman-$\alpha$ line of hydrogen \citep{vid03, lec10, kul14, ehr15}, metastable helium lines in the near-infrared \citep{spa18, nor18}, the helium triplet line at 1083~nm \citep{okl18,nor18,spa18}, near-ultraviolet observations \citep{sal19}, and soft X-ray measurements \citep{pop13}. 
The primary driver of atmospheric mass loss is thought to be the combined extreme-ultraviolet (EUV) and soft X-ray flux (XUV) emitted by the host star \citep{yel04, mur09}. 
While the EUV flux cannot be measured directly due to a lack of operating observatories in this wavelength range, it can be inferred from X-ray and UV measurements \citep{san11, fra13,joh21}. 

Quantifying stellar high-energy irradiation is essential for modeling exoplanetary atmospheric escape. 
However, significant uncertainties remain regarding the absorption height of X-rays in planetary atmospheres, the efficiency of mass loss, and the overall impact of high-energy radiation on planet formation \citep{owe14, coh15, don17, mon19}. 
X-ray observations, therefore, provide a critical, directly measurable constraint on one of the most influential factors driving atmospheric evaporation.

The 2019 launch of eROSITA has enabled the first comprehensive all-sky survey in X-rays since the ROSAT mission, offering an unprecedented dataset for studying stellar X-ray emissions \citep{bru22,mag22, car23,fre24}. 
Using SRG/eROSITA survey data, it is possible to estimate the combined X-ray and EUV (XUV) flux incident on exoplanets. 
Such measurements allow for the calculation of exoplanetary evaporation rates and the identification of highly irradiated systems that are prime targets for follow-up observations across multiple wavelengths.

This paper studies habitable regions around SRG/eROSITA coronal sources based on X-ray flux measurements obtained by fitting the X-ray spectra. 
The structure of this paper is as follows.
Section~\ref{sec_dat} explains the creation of the SRG/eROSITA data sample, while Section~\ref{sec_erosita_fits} describes the X-ray spectral modeling. 
Section~\ref{sec_xray_to_euv} details the computation of EUV fluxes using the X-ray fluxes. 
The habitable zone X-ray irradiation computation is shown in Section~\ref{sec_hab_region}, while the mapping of local hazard zones is described in Section~\ref{sec_hab_region}. 
A discussion of the results is presented in Section~\ref{sec_dis}, and the conclusions are summarized in Section~\ref{sec_con}. 
Throughout this work, errors are quoted at the 1$\sigma$ confidence level unless otherwise stated.  

\section{{\it eROSITA} Galactic objects sample selection}\label{sec_dat}
The {\it eROSITA} X-ray telescope \citep{pre21}, aboard the {\it Spectrum R\"otgen Gamma} (SRG) observatory, conducted the most extensive all-sky survey in the soft X-ray energy range ($0.2-10$~keV).  
During the first all-sky survey (eRASS1), {\it eROSITA} detected over 1 million sources, with $80\%$ of these sources identified as AGNs \citep{mer24}. 
Due to the bilateral data rights agreement within the SRG mission, the German {\it eROSITA} consortium has access to roughly half of the sky, which forms the basis of the present work.
\citet{fre24} carried out a cross-match between the coronal eRASS1 sources and the \textit{Gaia} third data release \citep[DR3, ][]{gai21,gai23} to compile a catalog of eRASS1 stellar sources, complete with \textit{Gaia} distances. 
Using the HamStar identification method \citep{sch21,fre22}, they identified 137,500 coronal sources with at least one optical counterpart and documented the characteristics. 

The X-ray spectra for this catalog were generated in \citet{gat24} using the \textit{eROSITA} data analysis software \texttt{eSASS} (version 201125 with 010 processing), which comprises 8,231 eRASS1 sources.
We used the {\tt srctool} task to produce source and background spectra, along with response files for each source identified in \citet{fre24}. 
The standard data reduction process involved creating circular extraction regions with radii scaled according to the maximum likelihood (ML) count rate from the eRASS1 source catalog. 
Background regions were similarly constructed as annuli, with their sizes adjusted to the ML count rate. Data from the Telescope Modules (TMs) were combined for TM 1-4,6 and TM 5,7. 
The latter group was analyzed only for energies above one keV due to excess soft emission caused by optical light leakage in TM 5 and 7 \citep{pre21}. 

For this work we have applied additional filtering of the \citet{gat24} data.
We excluded sources flagged for optical loading (\texttt{FLAG\_OPT=True}), which are optically very bright \citep{fre24}. 
For these sources, the X-ray detection is likely affected by the accumulation of optical/UV photons within a CCD pixel over the frame integration time. 
We also excluded those sources flagged as {\tt CORONAL=False}, as their X-ray emission is unlikely to be coronal but rather produced by an accretion process on a compact object.  
We note that many late-type stars appearing above the main sequence in \textsc{hamstar} are not necessarily giants but are instead pre-main sequence objects or unresolved binaries, both of which are characteristically X-ray bright.

To identify main sequence stars in our sample, we first derived their absolute magnitudes using  
\begin{equation}
M_G = G_\mathrm{mag} - 5 \log_{10}\left(\frac{d}{10}\right),
\end{equation}
where $G_\mathrm{mag}$ is the apparent \textit{Gaia} $G$-band magnitude and $d$ is the distance in parsecs. 
We then compared the derived absolute magnitudes with empirical main-sequence values at the same color, which were interpolated from the dwarf sequence compiled by \citet{pec12,pec13}.
Stars were considered consistent with the main sequence if they lay within a chosen tolerance around the interpolated relation.  
Following \citet{fre24}, we adopted a threshold of 1 mag in $M_G$. 
From the absolute magnitudes we computed the bolometric luminosity $(L_{\mathrm{bol}})$.

\section{X-ray Spectral Analysis}\label{sec_erosita_fits}
\citet{gat24} fitted each spectrum with multiple models selected to phenomenologically represent the most commonly observed spectral shapes in astronomical sources. 
Specifically, they used the following models (using {\sc xspec} nomenclature):

\begin{itemize}
\item Model A: An absorbed power-law model ({\sc XSPEC}: {\tt tbabs*pow}).
\item Model B: An absorbed thermal model ({\sc XSPEC}: {\tt tbabs*apec}).
\item Model C: An absorbed black-body model ({\sc XSPEC}: {\tt tbabs*bbody}).
\item Model D: An absorbed double thermal model ({\sc XSPEC}: {\tt tbabs*(apec+apec)}).
\item Model E: Same as Model D but with free abundances for the {\tt apec} components.
\end{itemize} 

Here, {\tt tbabs} represents the ISM X-ray absorption model as described in \citet{wil00}. 
By fitting the curvature of the X-ray spectra with these models, they estimated the column density ($N_{\mathrm{H}}$) values. 
The best-fit parameters, fit statistics, and statistical uncertainties are provided for each model. 
In this work, we used the unabsorbed fluxes from the best-fit model for each of these sources to investigate the effects of the coronal sources on their surrounding environment. 
As indicated in \citet{gat24}, the best fit is consistently obtained with models D and E, which provides a more robust physical interpretation of the source X-ray luminosities. 
Finally, it is important to note that the analysis by \citet{gat24} provides the most accurate X-ray fluxes, as it directly accounts for Galactic absorption along the line of sight. 
In contrast, 21-cm maps provide measurements over the entire Galactic line of sight and do not include the molecular contribution, potentially leading to an overestimation of column densities. 
 
\begin{figure}
    \centering
    \includegraphics[width=0.48\textwidth]{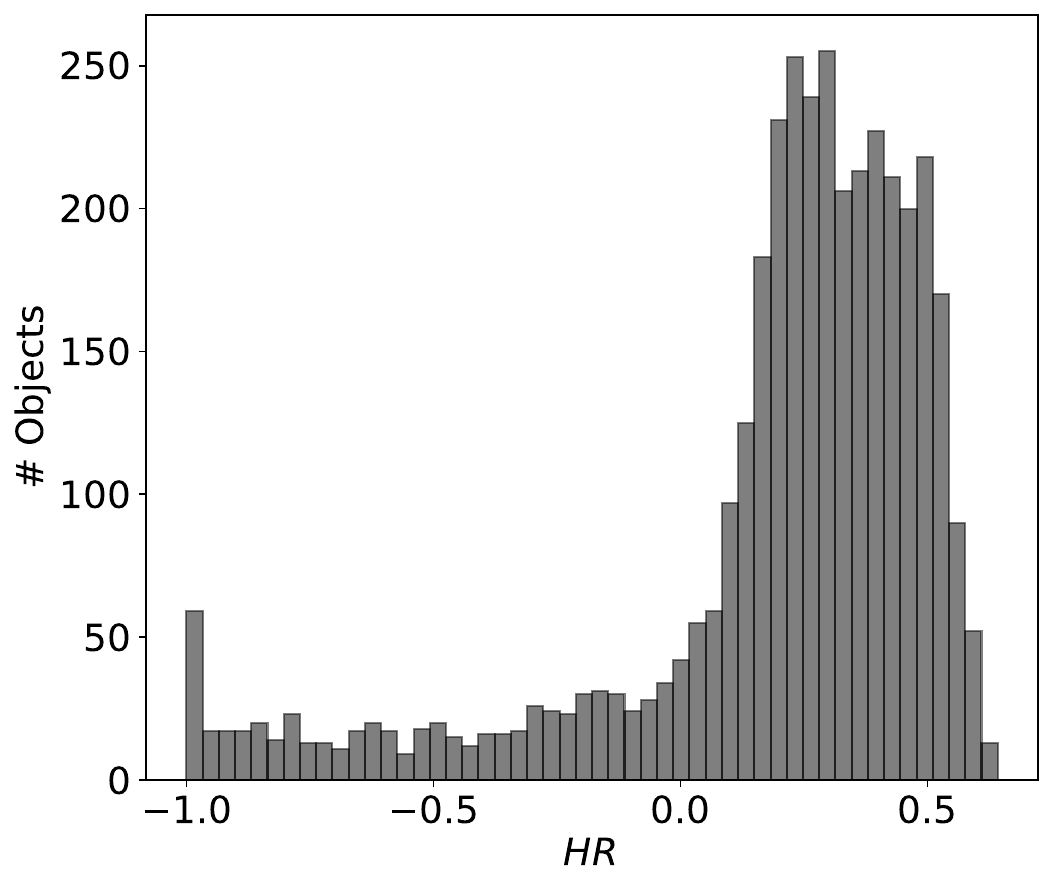}
    \caption{Distribution of hardness ratios for the analyzed sample. 
    The hardness ratio ($HR$) is defined as $HR = (H-S)/(H+S)$, where $H$ is the unabsorbed flux in the hard X-ray band ($0.5-1.0$~keV) and $S$ is the unabsorbed flux in the soft X-ray band ($0.2-0.5$~keV).
    The hardness ratio provides a proxy for the spectral shape, with higher values corresponding to harder spectra and lower values to softer emission.}
    \label{fig_his_hr}
\end{figure}

Figure~\ref{fig_his_hr} shows the Hardness Ratio distribution of the sources. 
The hardness ratio ($HR$) is defined as $HR = (H-S)/(H+S)$, where $H$ is the unabsorbed flux in the hard X-ray band ($0.5-1.0$~keV) and $S$ is the unabsorbed flux in the soft X-ray band ($0.2-0.5$~keV). 
The distribution shows that the majority of sources lie in the $HR \sim 0-0.5$ range, indicating spectra with significant flux in both the soft and hard bands. 
This pattern reflects that the sample contains typical coronal sources with hard X-ray emission. 

We computed X-ray luminosities for our sample by combining the unabsorbed fluxes ($F_{X}$) obtained from the spectral fits with the distances ($r$) derived from \textit{Gaia} parallaxes following the standard relation
\begin{equation}
L_{X} = 4 \pi r^2 F_{X}
\end{equation} 
This approach accounts for the intrinsic stellar brightness in X-rays by correcting for line-of-sight absorption, thus providing a uniform estimate of the coronal emission across the sample. 

To ensure physically meaningful XUV flux estimates and to remove spurious or non-stellar entries, we applied a conservative luminosity filter to the measured X-ray luminosities. 
Specifically, we retained only sources with $10^{26} \leq L_X \leq 10^{32}$~erg~s$^{-1}$, which encompasses the expected range of coronal X-ray emission for main-sequence stars, from quiet solar-like levels up to very active, saturated coronae. 
The lower bound excludes values consistent with non-detections, numerical artifacts, or incorrectly scaled fluxes, while the upper bound removes objects whose reported X-ray/XUV luminosities exceed plausible coronal emission (often a result of unit mismatches, anomalous fits, or erroneous distances). 
The final sample consists of 3750 main-sequence stars. 
All quoted maps and statistics use the filtered sample unless otherwise noted. 

\begin{figure}
       \centering 
\includegraphics[width=0.48\textwidth]{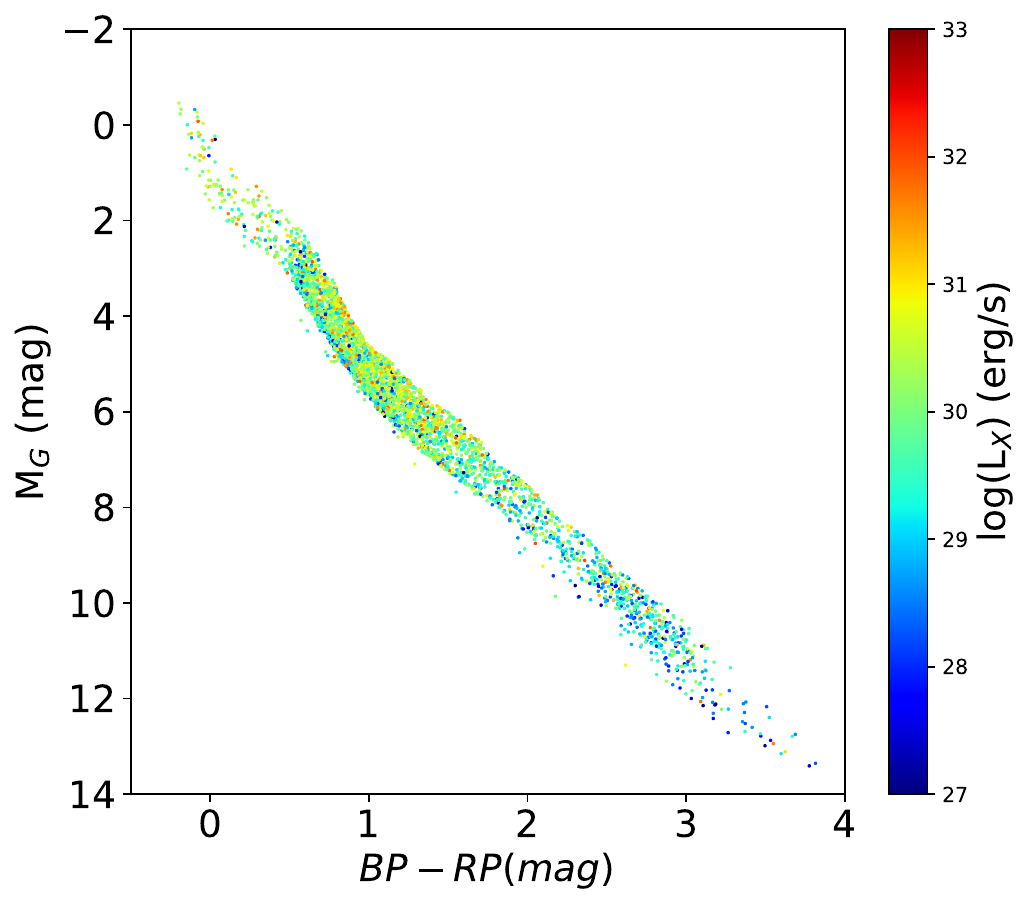} 
      \caption{
 Color-magnitude diagram of the sources, using absolute magnitude ($M_G$) and color ($BP-RP$) values from \textit{Gaia} DR3. 
 The color scales with X-ray luminosity ($L_{X}$), showing the relationship between stellar evolutionary status and coronal activity.
      }\label{fig_lx_bp_rp}
   \end{figure}

Figure~\ref{fig_lx_bp_rp} shows a \textit{Gaia} color-magnitude diagram of our sample, with points color-coded by their X-ray luminosity ($L_{X}$), which illustrates the relationship between stellar evolutionary status and coronal activity. 
Most sources lie along the main sequence, consistent with their coronal nature. 
The color coding highlights that hotter, more massive stars generally exhibit higher X-ray luminosities, while cooler, lower-mass stars span a broader range of $L_{X}$. 

\section{From X-rays to EUV}\label{sec_xray_to_euv}   
We investigated the impact of X-ray emissions on the habitability of planets, following the work done by \citet{mur09,kub20,han24,pop24}.
First, we computed the EUV luminosity ($L_{EUV}$) in the $0.013-0.1$~keV energy range by using the scaling relation from \citet{san11}:
\begin{equation}
\log{L_{EUV}}=(4.80\pm 1.99)+(0.860\pm 0.073)\log{L_{X}}
\end{equation}
where $L_{X}$ is the X-ray luminosity, in erg s$^{-1}$ and in the $0.1-2.4$~keV energy range, obtained from the best-fit spectral model in ergs~s$^{-1}$.
This scaling relation is valid only for main-sequence stars, thus justifying our sample selection process (see Section~\ref{sec_dat}). 

\begin{figure}
\centering
\includegraphics[width=0.48\textwidth]{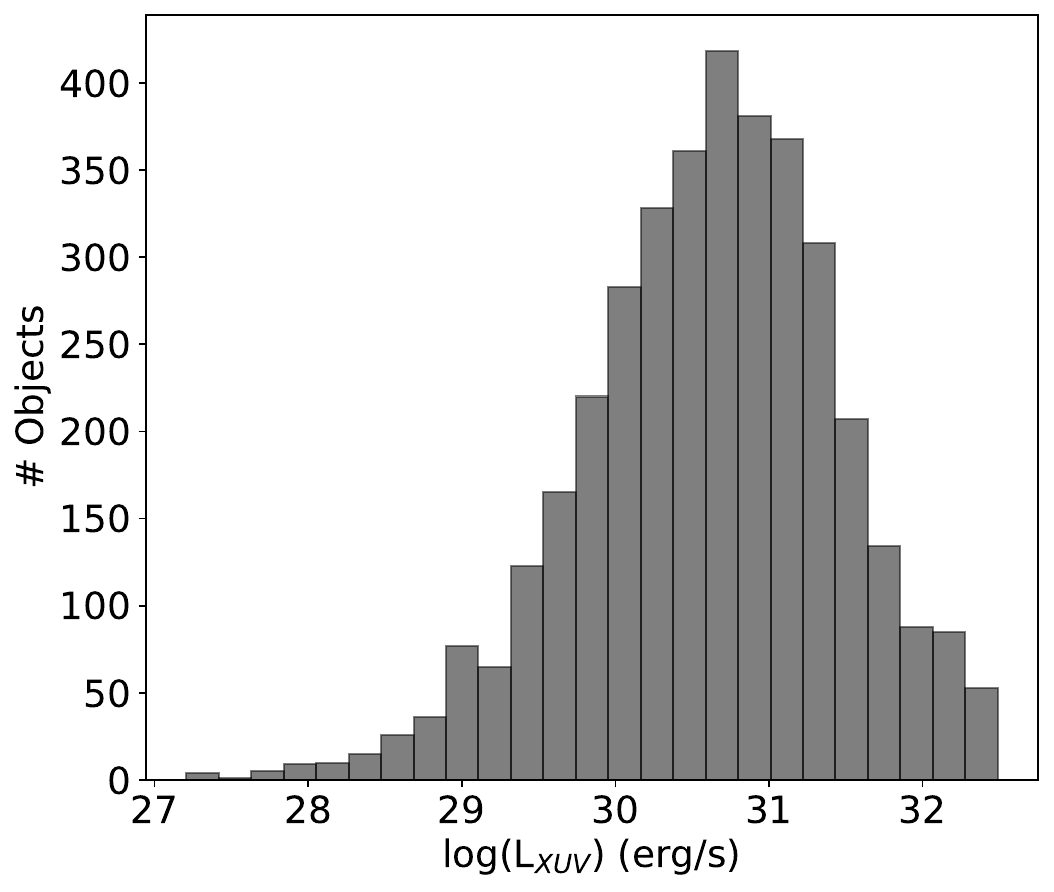}
\caption{
Histogram of the combined X-ray and extreme-UV luminosity ($L_{\mathrm{XUV}}$) for the main-sequence sample.  
}
\label{fig_hist_l_xuv}
\end{figure}

\begin{figure}
\centering
\includegraphics[width=0.48\textwidth]{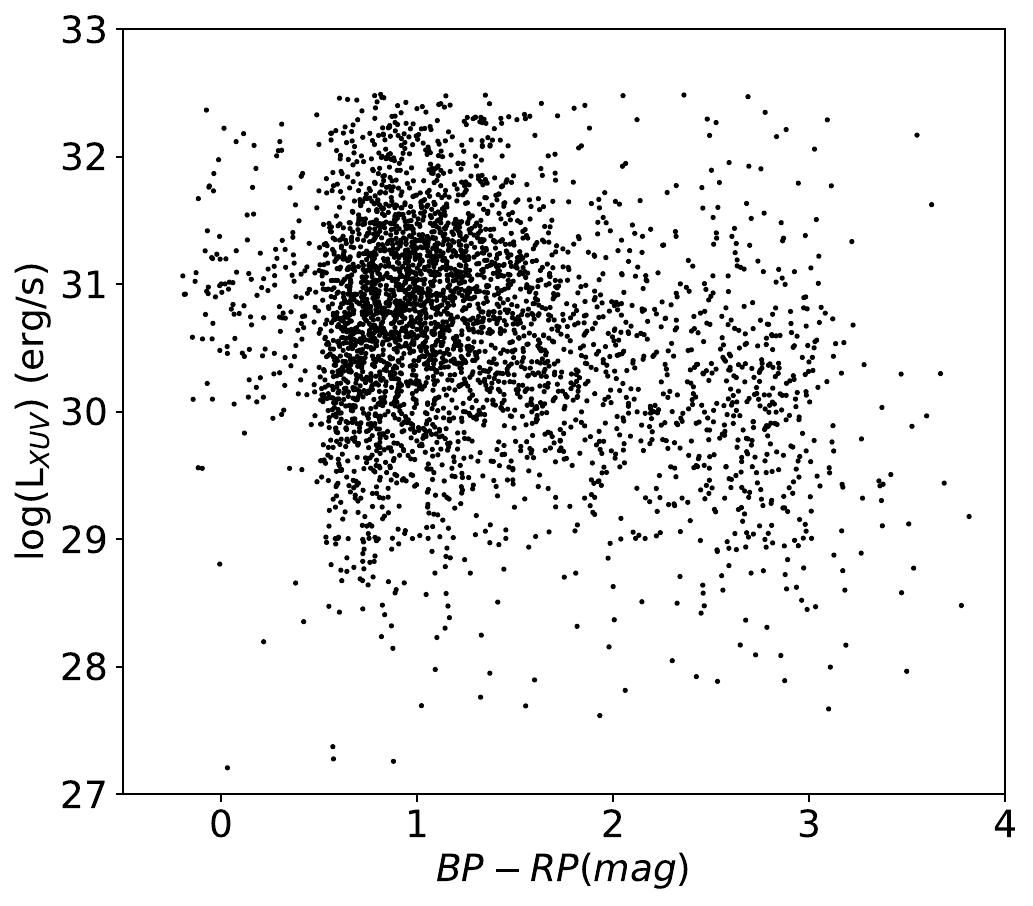}
\caption{
Combined X-ray and extreme-UV luminosity ($L_{\mathrm{XUV}}$) as a function of \textit{Gaia} $BP-RP$ color. 
Only sources with $\log_{10}(L_{\mathrm{XUV}})$ between $10^{27}$ and $10^{33}$~erg~s$^{-1}$ are shown to improve visualization.
}
\label{fig_l_xuv_bp_rp}
\end{figure}

\begin{figure}
\centering
\includegraphics[width=0.48\textwidth]{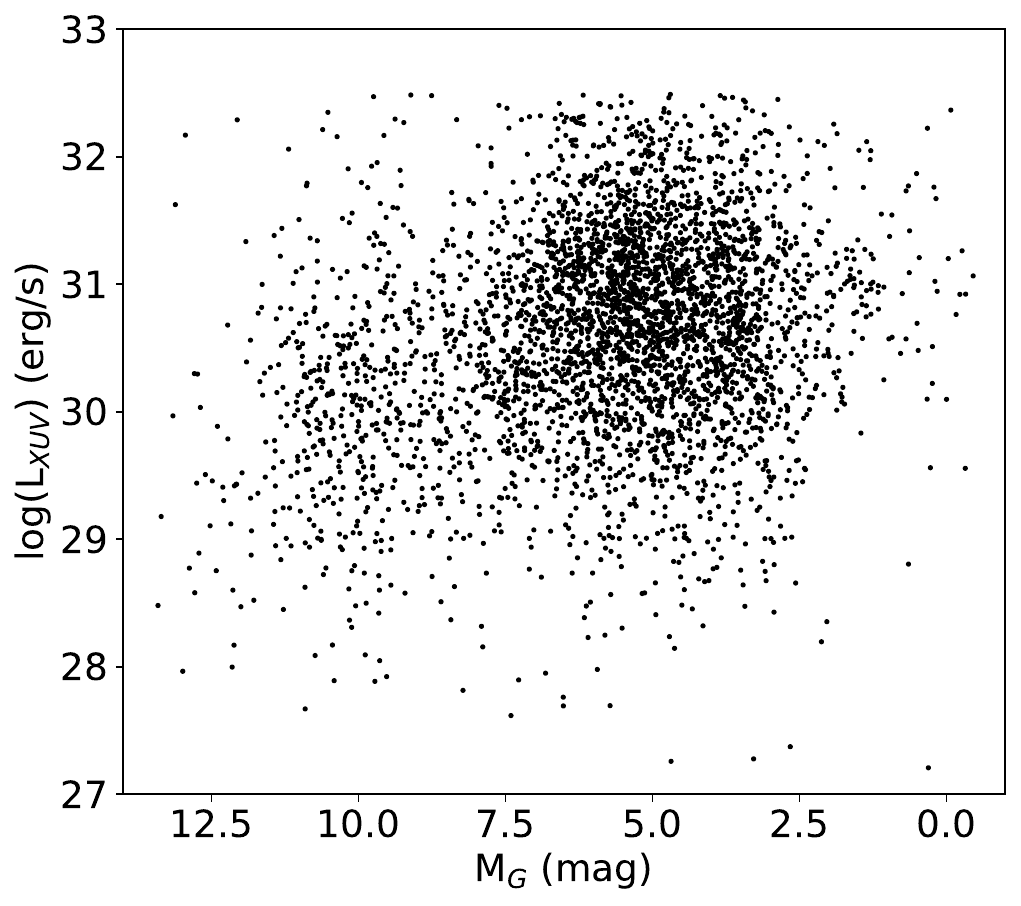}
\caption{
Combined X-ray and extreme-UV luminosity ($L_{\mathrm{XUV}}$) as a function of \textit{Gaia} absolute magnitude $M_G$.  
}
\label{fig_l_xuv_mg}
\end{figure}

\begin{figure}
\centering
\includegraphics[width=0.48\textwidth]{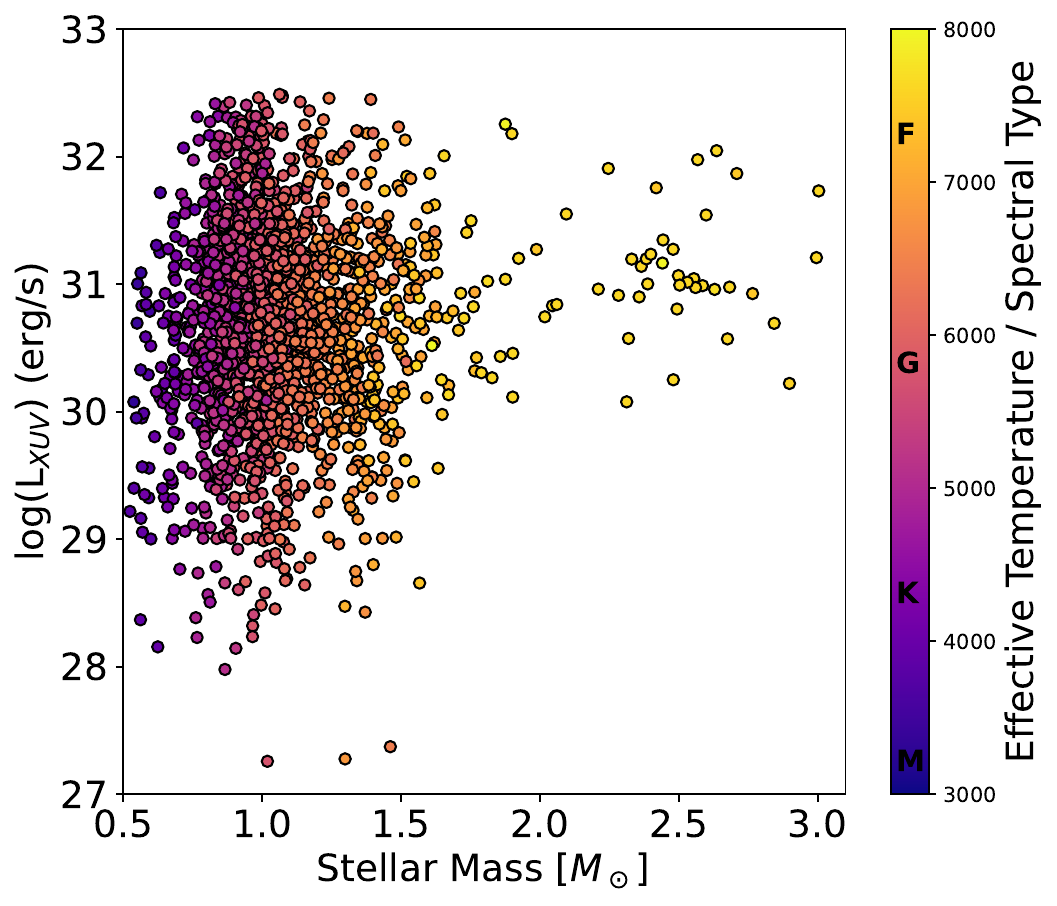}
\caption{
Combined X-ray and extreme-UV luminosity ($L_{\mathrm{XUV}}$) as a function of stellar mass. 
The EUV component is derived using the \citet{san11} scaling relation. 
The spectral type and temperatures are also indicated.
}
\label{fig_l_xuv_mass}
\end{figure}

Figures~\ref{fig_hist_l_xuv}--\ref{fig_l_xuv_mass} provide complementary views of the high-energy emission of our main-sequence sample.
Figure~\ref{fig_hist_l_xuv} presents a histogram of the combined X-ray and EUV (XUV) luminosity ($L_{\mathrm{XUV}}=\log{L_{EUV}}+L_{X}$), highlighting the distribution and typical values of high-energy luminosities. 
Figure~\ref{fig_l_xuv_bp_rp} shows $L_{\mathrm{XUV}}$ as a function of \textit{Gaia} $BP-RP$ color. 
Figure~\ref{fig_l_xuv_mg} presents the relationship between $L_{\mathrm{XUV}}$ and absolute magnitude ($M_G$).
Finally, Figure~\ref{fig_l_xuv_mass} illustrates the dependence of $L_{\rm XUV}$ on stellar mass, showing that more massive stars are typically more XUV luminous. 
These figures collectively link fundamental stellar properties with coronal activity, providing a basis for assessing the potential high-energy environment within the habitable zones of different stellar types. 
   
\section{Habitable Zone X-ray Irradiation}\label{sec_hab_region}  
\begin{figure}
\centering
\includegraphics[width=0.48\textwidth]{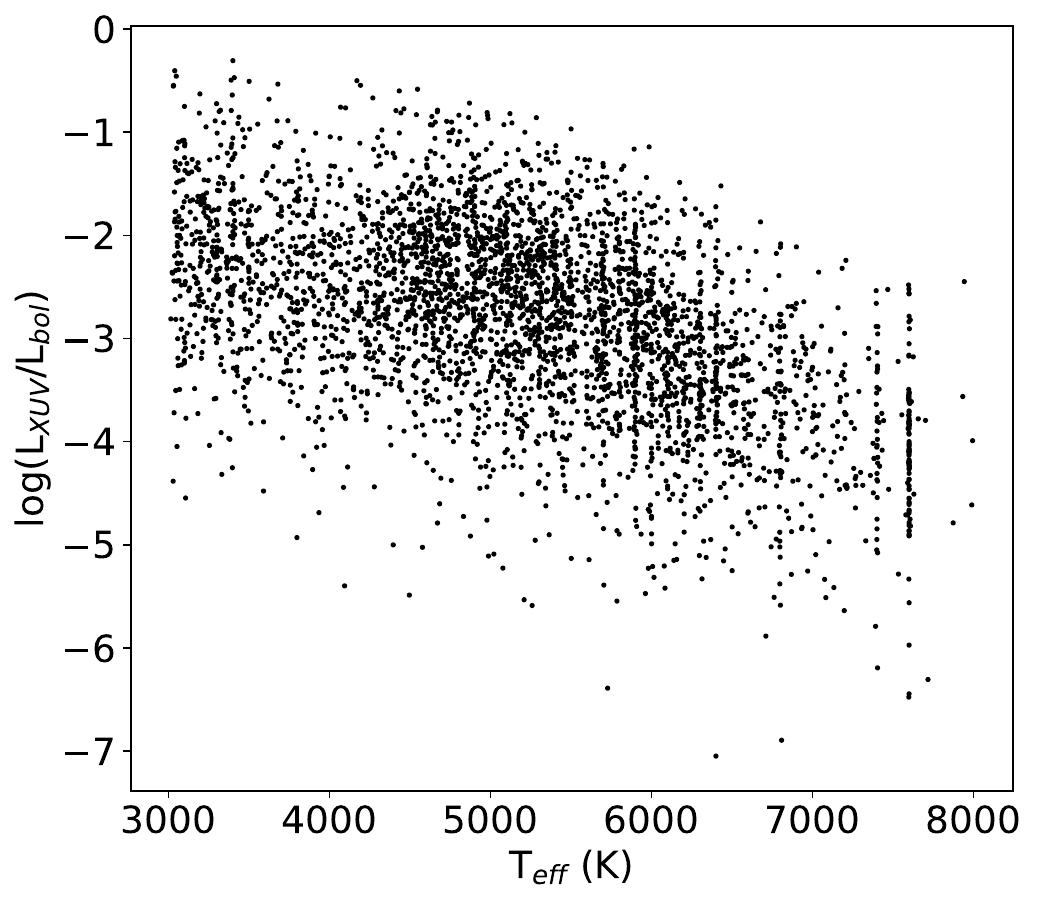}
\caption{
Ratio of combined X-ray and extreme-UV luminosity to bolometric luminosity, $L_{\mathrm{XUV}}/L_{\mathrm{bol}}$, as a function of stellar effective temperature, $T_{\mathrm{eff}}$. 
}
\label{fig_Lxuv_Lbol_teff}
\end{figure}

Figure~\ref{fig_Lxuv_Lbol_teff} presents the ratio of combined X-ray and extreme-UV (XUV) luminosity to bolometric luminosity, $L_{\mathrm{XUV}}/L_{\mathrm{bol}}$, as a function of stellar effective temperature, $T_{\mathrm{eff}}$.  
This plot reveals the dependence of stellar high-energy emission on temperature: cooler stars, typically more magnetically active, show higher $L_{\mathrm{XUV}}/L_{\mathrm{bol}}$ ratios, whereas hotter stars generally emit a smaller fraction of their total luminosity in the X-ray and EUV bands. 
Most stars in our sample exhibit $L_{\mathrm{XUV}}/L_{\mathrm{bol}} \ll 1$, which indicates that only a small fraction of their total energy is emitted in the high-energy bands, as expected for main-sequence stars.

To estimate the high-energy irradiation that a hypothetical planet would receive in the habitable zone (HZ) of each star, we first computed the effective stellar flux defining the HZ following \citet{kop14}. 
The effective flux ($S_{\mathrm{eff}}$) is a function of the stellar effective temperature and can be expressed as a fourth-order polynomial relative to the solar effective temperature:
\begin{equation}
S_{\mathrm{eff}} = S_{\mathrm{eff},\odot} + a\,T_{\star} + b\,T_{\star}^2 + c\,T_{\star}^3 + d\,T_{\star}^4,
\end{equation}
where $T_{\star} = T_{\mathrm{eff}} - 5780$~K and the coefficients $S_{\mathrm{eff},\odot}, a, b, c, d$ correspond to the inner or outer edges of the HZ depending on the chosen model. 
The distance of the HZ is then obtained as
\begin{equation}
d_{\mathrm{HZ}} = \sqrt{\frac{L_{\mathrm{bol}}/L_{\odot}}{S_{\mathrm{eff}}}}\ \mathrm{AU},
\end{equation}
where $L_{\mathrm{bol}}$ is the stellar bolometric luminosity and $L_{\odot}$ is the solar bolometric luminosity. 
Finally, the X-ray and extreme-UV (XUV) flux at the habitable zone is calculated as
\begin{equation}
F_{\mathrm{XUV,HZ}} = \frac{L_{\mathrm{XUV}}}{4\pi d_{\mathrm{HZ}}^2},
\end{equation}
where $L_{\mathrm{XUV}}$ is the combined stellar X-ray and EUV luminosity. 
The parameters used in the model are listed in Table~\ref{tab_seff_par}, and are obtained from \citet{kop14}. 

\begin{table} 
\caption{\label{tab_seff_par}Effective flux parameters from \citet{kop14}.}
\centering
\begin{tabular}{lc}
\hline
Parameter & Value \\
\hline 
$S_{\mathrm{eff},\odot}$ & $1.107$ \\  
$a$ & $1.332\times 10^{-4}$ \\    
$b$ & $1.580\times 10^{-8}$ \\    
$c$ & $-8.308\times 10^{-12}$ \\    
$d$ & $-1.931\times 10^{-15}$ \\    
\hline    
\end{tabular}
\end{table}

\begin{figure}
\centering
\includegraphics[width=0.48\textwidth]{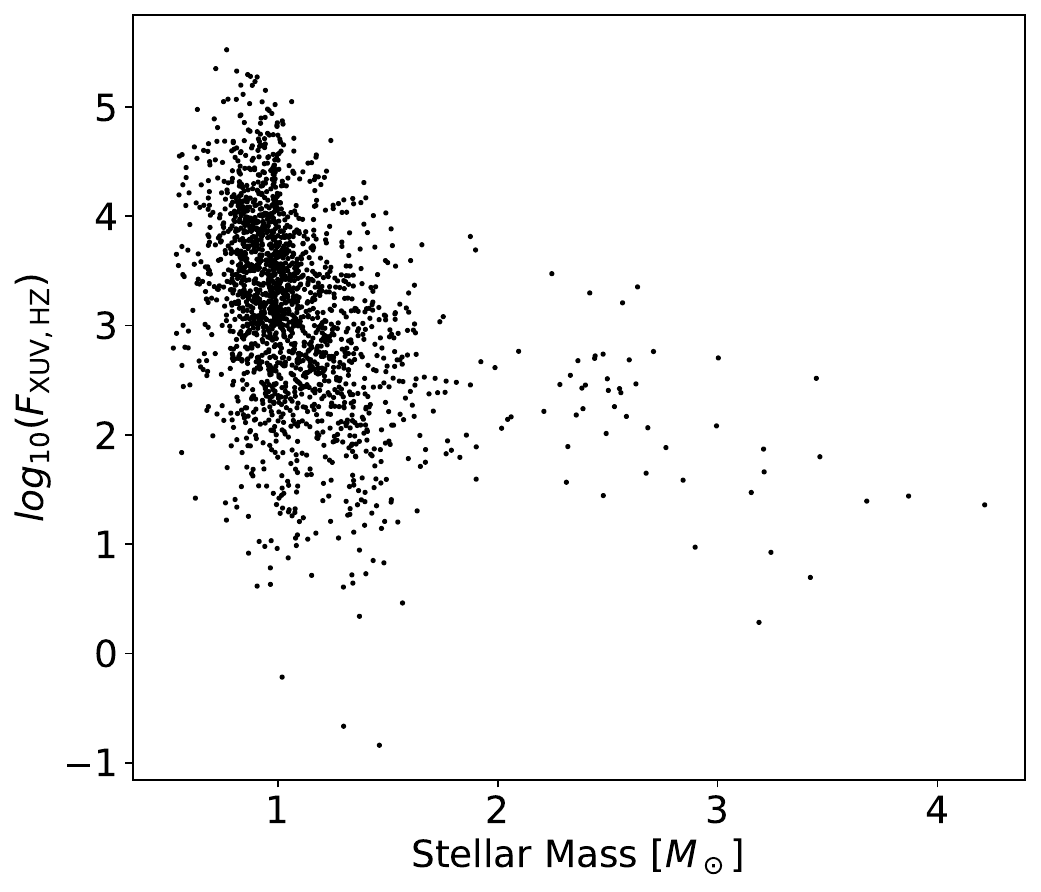}
\caption{
X-ray plus EUV flux ($F_{\mathrm{XUV,HZ}}$) received at the habitable zone of each star as a function of stellar mass. Only sources with $\log_{10}(F_{\mathrm{XUV,HZ}})$ between $-1$ and $6$ are shown to improve visualization.
}
\label{fig_fxuv_hz_mstar}
\end{figure}

Figure~\ref{fig_fxuv_hz_mstar} shows the high-energy flux that a planet would receive at the center of the habitable zone as a function of stellar mass (provided by \textit{Gaia}). 
The plot illustrates that low-mass stars tend to expose their habitable zones to higher XUV fluxes due to their closer-in habitable zones. 
Conversely, more massive stars, despite having higher intrinsic XUV luminosities, have more distant habitable zones, which results in lower fluxes at the HZ. 
The majority of the computed XUV fluxes in the habitable zones fall between $10^{0}$ and $10^{5}$~erg~cm$^{-2}$~s$^{-1}$, which is significantly larger than the present-day solar XUV flux at Earth ($\sim 4$~erg~cm$^{-2}$~s$^{-1}$) and indicates potentially XUV-hazardous environments.
The extremes correspond to very low-activity stars or highly luminous XUV emitters.

\subsection{Mass-loss rate of exoplanets}\label{sec_mass_loss_rate} 
For a given planet mass and radius located within the continuous habitable zone ($d_{\mathrm{HZ}}$), the photoevaporative mass loss rate ($\dot{m}$) due to high-energy stellar emission can be computed as \citep[see e.g.,][]{owe12,lop12}:
\begin{equation} 
\dot{m}=\eta \frac{3\beta^{3}F_{XUV}}{4GK\rho_{pl}}
\end{equation}  

where $\rho_{pl}$ is the density of the planet, $\beta$ is the effective XUV absorption cross-section, $\eta$ is the efficiency of the atmospheric escape (assumed to be 0.15), and $G$ is the gravitational constant.

The factor $K$ accounts for the effect of Roche lobe overflows, as described by \citet{erk07} as:
\begin{equation}
K=1-\frac{3}{2\xi}+\frac{1}{2\xi^{3}}
\end{equation} 
and 
\begin{equation}
\xi=\left(\frac{M_{pl}}{3M_{*}}\right)^{1/3}\frac{a}{R_{pl}}
\end{equation}
with $M_{pl}$ as the mass of the planet, $M_{*}$ as the mass of the host star, $a$ as the orbital distance between the star and the planet (taken as $d_{\mathrm{HZ}}$), and $R_{pl}$ as the radius of the planet.

\begin{figure}
\centering 
\includegraphics[width=0.47\textwidth]{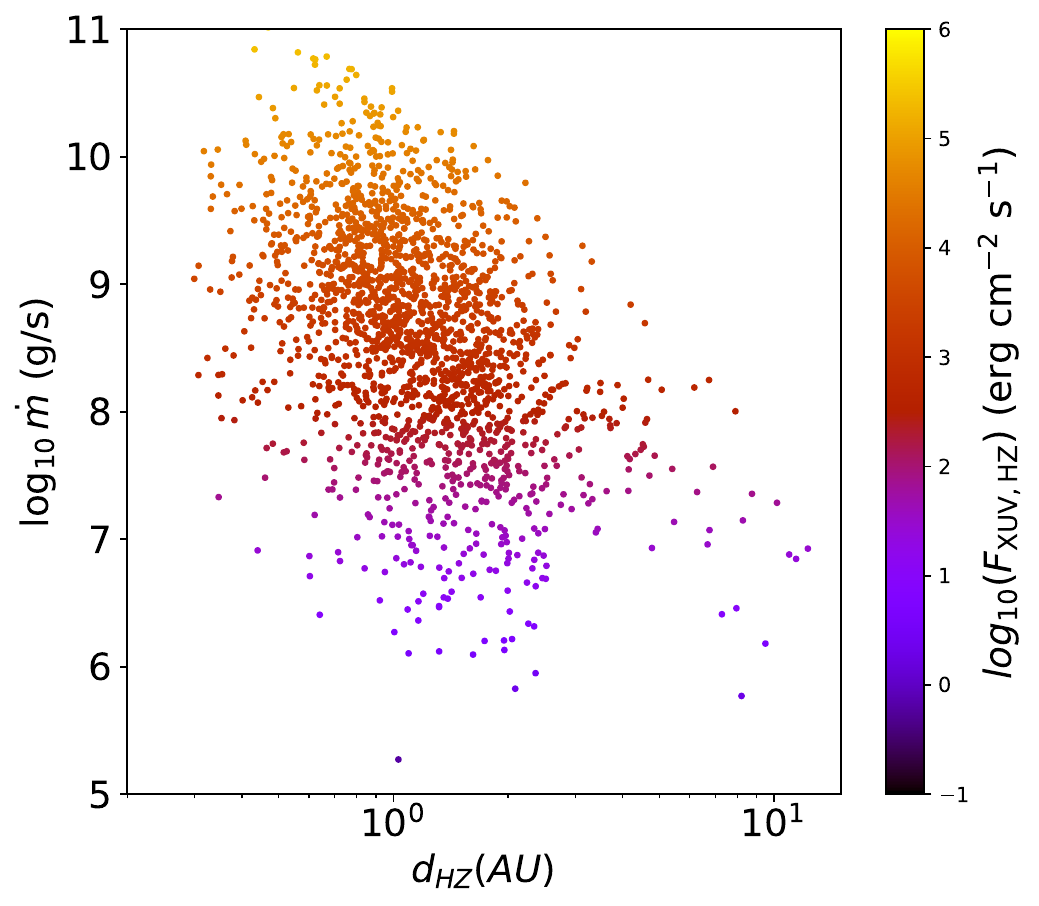} 
\caption{
Calculated photoevaporative mass-loss rates for a theoretical Earth-like planet ($M_{pl}=5.972\times 10^{27}$~g, $R_{pl}=6.378\times 10^{8}$~cm) orbiting the sample of stars. 
Each data point represents a unique star-planet combination. 
The mass-loss rate ($\log_{10}\dot{m}$) is plotted against the host star's Habitable Zone distance ($d_{\mathrm{HZ}}$). 
The color scale indicates the incident X-ray and extreme-UV (XUV) flux at the habitable zone ($F_{\mathrm{XUV,HZ}}$). 
}
\label{fig_mass_dhz}
\end{figure}

Figure~\ref{fig_mass_dhz} shows the estimated atmospheric mass-loss rates for a hypothetical terrestrial planet ($M_{pl}=5.972\times 10^{27}$~g, $R_{pl}=6.378\times 10^{8}$~cm) as a function of the habitable zone distance ($d_{\mathrm{HZ}}$) for each star in our sample.  
The color-coding represents the incident X-ray and extreme-UV flux ($F_{\mathrm{XUV,HZ}}$). 
The dependence on $F_{\mathrm{XUV,HZ}}$ is particularly evident, with strongly irradiated systems producing the most extreme escape rates. 
These results highlight the critical role of stellar high-energy output in determining the long-term atmospheric retention and potential habitability of exoplanets. 
For comparison, observed escape rates in highly irradiated close-in exoplanets such as HD~209458b or GJ~436b are typically in the range of $10^{9}$--$10^{11}$~g~s$^{-1}$ \citep[e.g.,][]{vid03,ehr15,lec10}, consistent with our model predictions for highly irradiated systems.

\section{Mapping Local Hazard Zones}\label{sec_hab_region}  
\begin{figure*}
\centering
\includegraphics[width=0.48\textwidth]{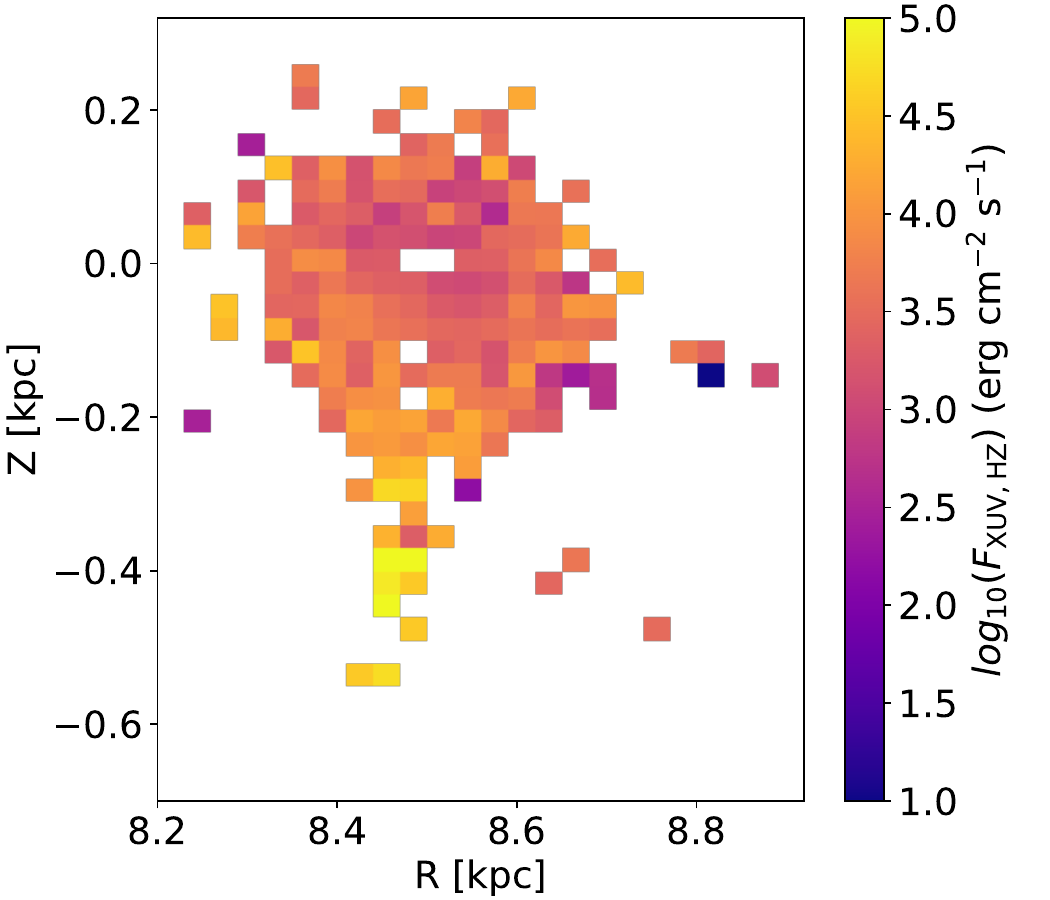}
\includegraphics[width=0.48\textwidth]{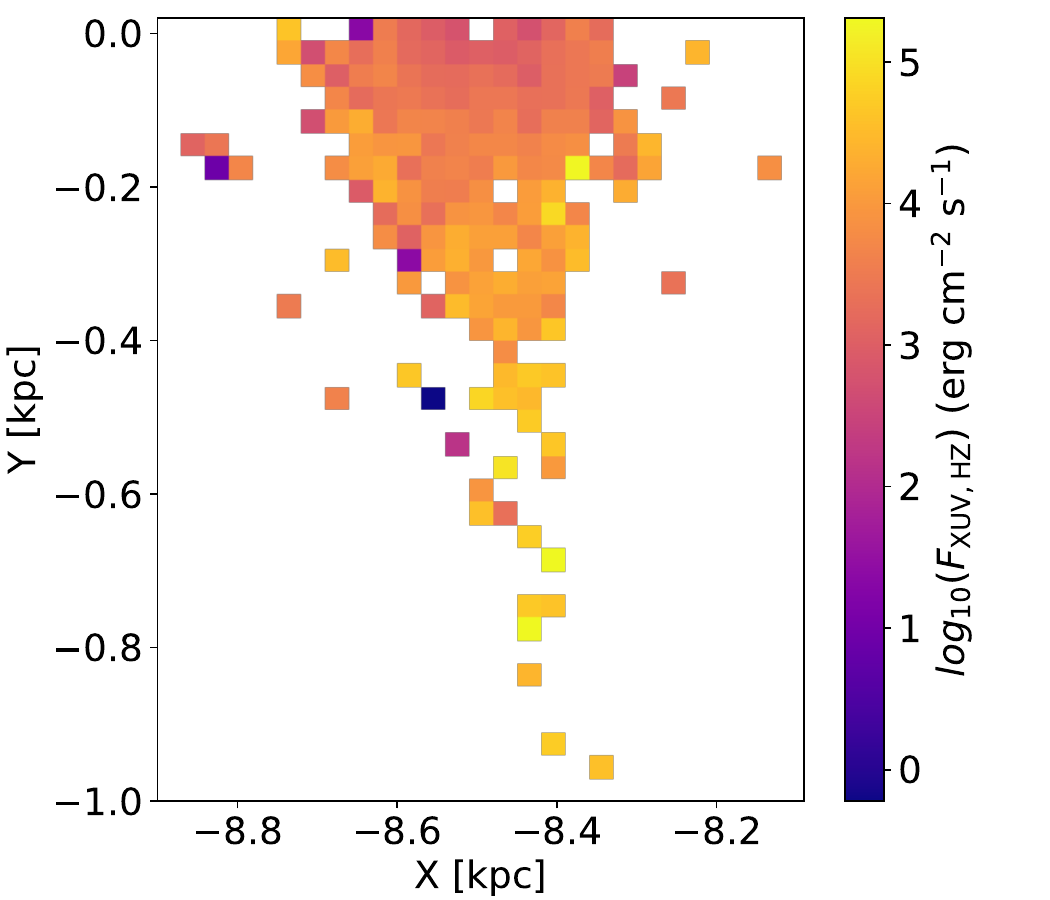} 
\caption{
        \textbf{Left panel:} Map of the average XUV flux in the habitable zone, $\langle F_{\mathrm{XUV,HZ}} \rangle$, in the Galactic R-Z plane. 
        \textbf{Right panel:} Map of the average XUV flux in the habitable zone, $\langle F_{\mathrm{XUV,HZ}} \rangle$, in the Galactic X-Y plane.
}
\label{fig_avg_flux_rz_xy}
\end{figure*}

Figure~\ref{fig_avg_flux_rz_xy} shows the map of the average XUV flux in the habitable zone ($\langle F_{\mathrm{XUV,HZ}} \rangle$) in the Galactic plane for our sample of stars. 
To construct these 2D maps, we binned the stellar sample in cylindrical coordinates, defined as $(R,\phi,Z)$ with
\begin{equation}
R = \sqrt{X^2 + Y^2}, \quad \phi = \arctan\!\left(\frac{Y}{X}\right), \quad Z = Z ,
\end{equation}
where $(X,Y,Z)$ are the Cartesian Galactocentric coordinates. We then computed the average XUV flux in each bin. 
This was done by weighting the average with $F_{\mathrm{XUV,HZ}}$ and dividing by the number of stars per bin. 
This approach accounts for the non-uniform stellar distribution and provides a representative measure of the typical irradiation environment in each region. 
The map was constructed using bins of $30$~pc in both the radial and vertical directions. 
This map illustrates how high-energy irradiation is distributed across the local Galactic environment, indicating the levels of XUV flux that hypothetical planets would receive in their habitable zones. 
Regions with higher average flux correspond to more X-ray active stellar neighborhoods, which could potentially affect atmospheric retention on orbiting planets. 

\begin{figure*}
\centering
\includegraphics[width=0.48\textwidth]{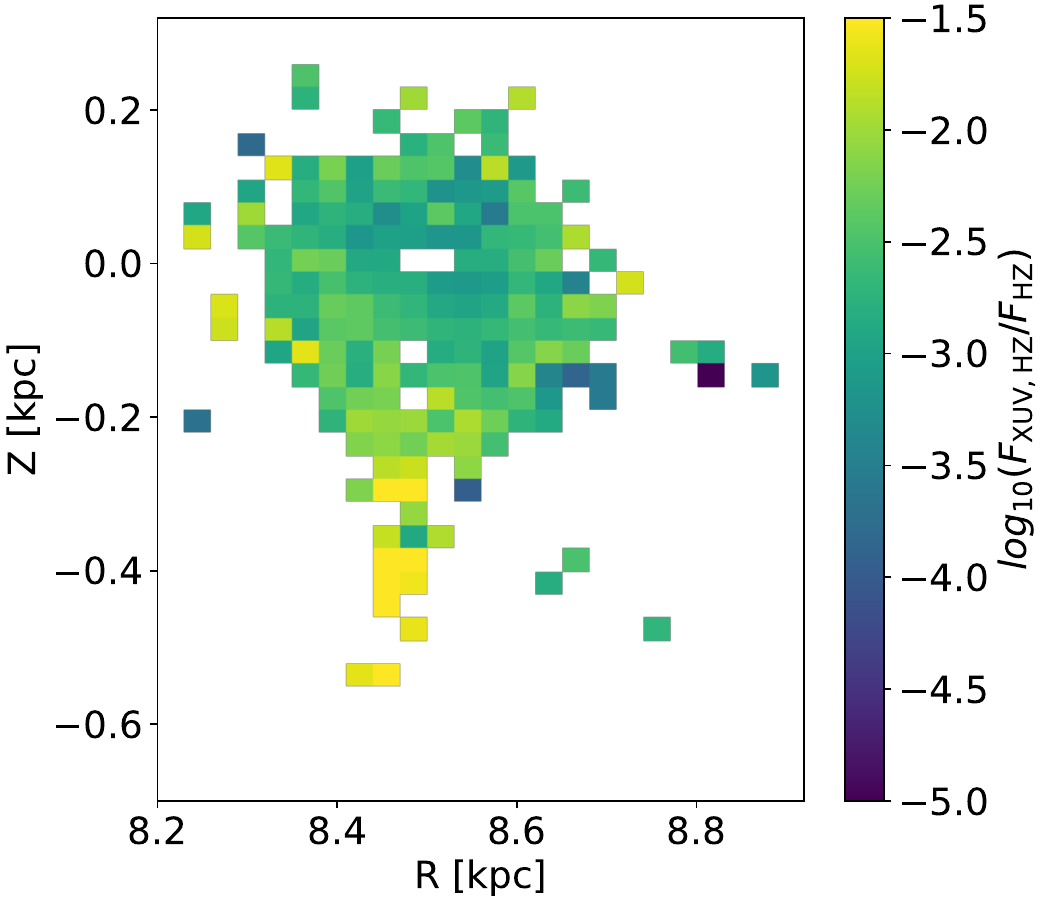}
\includegraphics[width=0.48\textwidth]{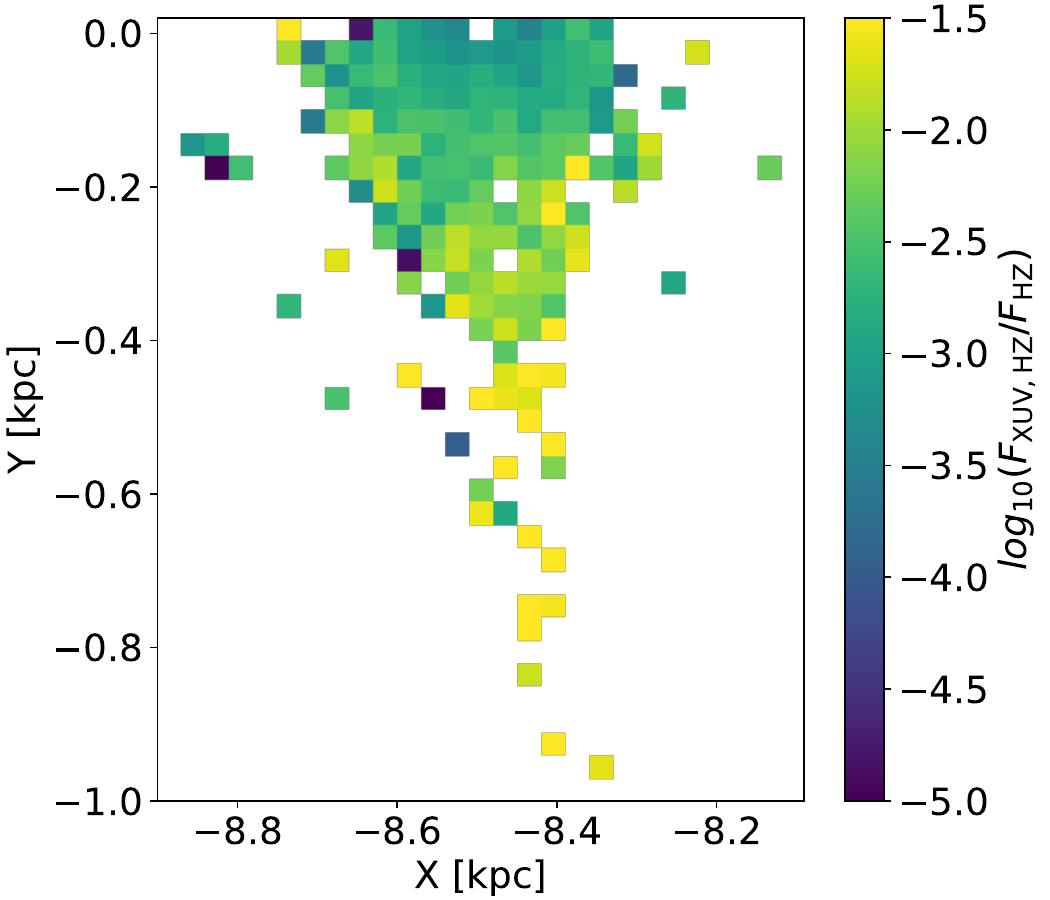}
\caption{
        \textbf{Left panel:} Map of the normalized XUV flux in the habitable zone, $F_{\mathrm{XUV,HZ}}/F_{\mathrm{HZ}}$, in the Galactic R-Z plane. 
        \textbf{Right panel:} Map of the normalized XUV flux in the habitable zone, $F_{\mathrm{XUV,HZ}}/F_{\mathrm{HZ}}$, in the Galactic X-Y plane.
}
\label{fig_ratio_flux_rz_xy}
\end{figure*}

Figure~\ref{fig_ratio_flux_rz_xy} presents the normalized XUV flux ratio, $F_{\mathrm{XUV,HZ}}/F_{\mathrm{HZ}}$, in the Galactic plane.  
Most values are clustered around $0.01-0.1$, while the plotted range is chosen to enhance visualization. 
This ratio quantifies the fraction of stellar high-energy emission relative to the total bolometric flux at the habitable zone. 
Areas with higher ratios indicate environments where XUV irradiation is proportionally stronger, suggesting potentially harsher conditions for planetary atmospheres. 
These maps provide a Galactic-scale view of ``hazard zones,'' identifying regions where high-energy stellar radiation could significantly influence habitability.
For reference, the Sun XUV flux in its habitable zone is $F_{\rm XUV,HZ,\odot} \sim 4~{\rm erg~cm^{-2}~s^{-1}}$, while the bolometric flux at $1$~AU is $F_{\rm HZ,\odot} \sim 1.36\times10^6~{\rm erg~cm^{-2}~s^{-1}}$, giving a ratio $F_{\rm XUV,HZ}/F_{\rm HZ} \sim 3\times10^{-6}$. 
This indicates that most stars in our sample are significantly more XUV-active than the Sun.

\section{Discussion}\label{sec_dis}
Our study provides an estimation of atmospheric mass-loss rates for exoplanets, focusing on the energy-limited escape mechanism. 
Our results are broadly consistent with previous studies, confirming that the energy-limited approximation is a useful tool for estimating mass-loss rates in many cases. 
\citet{ehr11} estimated mass-loss rates for transiting exoplanets, finding values ranging from $10^6$ to $10^{13}$~g~s$^{-1}$ under various assumptions.
Similarly, \citet{kre21} critically assessed the applicability of this approximation, concluding that it provides correct order-of-magnitude estimates for approximately $76\%$ of planets.
However, for planets with very low or very high gravitational potentials, or those subject to high equilibrium temperatures and strong irradiation fluxes, the energy-limited approximation can deviate from hydrodynamic simulation predictions by up to three orders of magnitude in individual cases \citep[e.g.,][]{kub18,kre21,cal22}.
This work further highlights the complex relationship between the incident X-ray flux on the planet, and its atmospheric mass-loss rate.   

Understanding atmospheric mass loss is crucial for modeling the long-term evolution of exoplanetary atmospheres. 
Previous works \citep[e.g.,][]{san11,owe12,lop14} have shown that XUV-driven mass loss can significantly alter the atmospheric composition and survival of close-in planets around active stars. 
\citet{fos22} conducted a comprehensive study utilizing \textit{eROSITA} data to estimate X-ray and EUV irradiation fluxes and corresponding mass-loss rates for 287 known exoplanets. 
 
Our findings build upon this body of work by providing updated mass-loss rates for a large sample of main-sequence stars, using an extensive catalog of coronal sources. 
We stress that our sample is flux-limited and thus biased towards the most X-ray luminous stars, which naturally results in higher average XUV fluxes compared to a solar-like star. 
For reference, the present-day XUV flux at the solar habitable zone is $\sim 4$~erg~cm$^{-2}$~s$^{-1}$, while our sample exhibits values ranging from $10^0$ to $10^5$~erg~cm$^{-2}$~s$^{-1}$. 
It is important to note that coronal activity decreases with stellar age \citep{rib05,tu15,joh21}, and most of the sources in our sample are likely young, magnetically active stars. 
At comparable ages, the young Sun is also believed to have emitted significantly stronger XUV radiation than today, by up to two orders of magnitude \citep{rib05,cla12}. 
This evolutionary context suggests that our high flux levels are consistent with expectations for young stellar populations and highlight the temporal dependence of atmospheric erosion processes. 
The maps we present therefore provide a first-order view of the local distribution of potentially hazardous radiation environments, illustrating where exoplanetary atmospheres may be most strongly affected by stellar XUV irradiation. 

Our analysis also provides novel insights into the characteristics of the stellar sample itself. 
We find that the ratio of XUV to bolometric luminosity ($L_{\mathrm{XUV}}/L_{\mathrm{bol}}$) is generally higher for cooler, more magnetically active stars, which indicates that they pose a greater potential hazard to planetary atmospheres. 
The maps of average XUV flux and normalized XUV flux ratio in the Galactic plane provide a Galactic-scale view of ``hazard zones,'' which can be used to identify regions where high-energy stellar radiation could significantly influence planetary habitability.

While the energy-limited model provides a valuable first-order estimate, it does not account for detailed hydrodynamic processes, magnetic fields, or core-powered mass-loss mechanisms, which can become significant in extreme irradiation environments \citep{owe16,tho18}. 
Future work should incorporate these effects to improve the accuracy of mass-loss predictions, particularly for hot Jupiters and highly irradiated super-Earths. 
The unprecedented data from \textit{eROSITA} provides a critical foundation for these future studies, offering a robust dataset to constrain stellar XUV fluxes and refine our understanding of exoplanetary atmospheric escape.

\section{Conclusions}\label{sec_con}
In this work, we have utilized the extensive all-sky X-ray survey data from \textit{SRG/eROSITA} to study stellar high-energy radiation and its impact on exoplanet habitability. 
By combining our robust sample of 3750 main-sequence stars with \textit{Gaia} astrometry, we computed XUV luminosities and fluxes at the habitable zone. 
Our analysis revealed that the majority of stars in our sample are significantly more XUV-active than the Sun, with habitable zone fluxes ranging from $10^0$ to $10^5$ erg~cm$^{-2}$~s$^{-1}$. 
We found that cooler, more active stars exhibit a higher ratio of XUV to bolometric luminosity ($L_{\mathrm{XUV}}/L_{\mathrm{bol}}$), and we created Galactic-scale maps identifying ``hazard zones'' where XUV irradiation is highest. 
We model atmospheric mass loss ratios for earth-like planets located in the habitable zone for each star, showing the importance of the incident X-ray flux in the photoevaporation process. 
These results highlight the critical role of stellar high-energy emission in planetary atmospheric evolution and emphasize the need for direct XUV observations to fully understand the long-term habitability of exoplanets.

\begin{acknowledgements} 
This work is based on data from {\it eROSITA}, the soft X-ray instrument aboard SRG, a joint Russian-German science mission supported by the Russian Space Agency (Roskosmos), in the interests of the Russian Academy of Sciences represented by its Space Research Institute (IKI), and the Deutsches Zentrum f\"ur Luft- und Raumfahrt (DLR). 
The SRG spacecraft was built by Lavochkin Association (NPOL) and its subcontractors, and is operated by NPOL with support from the Max Planck Institute for Extraterrestrial Physics (MPE). 
The development and construction of the {\it eROSITA} X-ray instrument was led by MPE, with contributions from the Dr. Karl Remeis Observatory Bamberg \& ECAP (FAU Erlangen-Nuernberg), the University of Hamburg Observatory, the Leibniz Institute for Astrophysics Potsdam (AIP), and the Institute for Astronomy and Astrophysics of the University of T\"ubingen, with the support of DLR and the Max Planck Society. 
The Argelander Institute for Astronomy of the University of Bonn and the Ludwig Maximilians Universit\"at Munich also participated in the science preparation for {\it eROSITA}.  
This research was carried out on the High Performance Computing resources of the cobra cluster at the Max Planck Computing and Data Facility (MPCDF) in Garching operated by the Max Planck Society (MPG). 
The {\it eROSITA} data shown here were processed using the eSASS software system developed by the German {\it eROSITA} consortium.    
This research was carried out on the High Performance Computing resources of the cobra cluster at the Max Planck Computing and Data Facility (MPCDF) in Garching operated by the Max Planck Society (MPG)
\end{acknowledgements}

\bibliographystyle{aa}

\end{document}